\input psfig

\catcode `\@=11 

\def\@version{1.3}
\def\@verdate{28.11.1992}


%
%
%
%
%
%

\font\fiverm=cmr5
\font\fivei=cmmi5	\skewchar\fivei='177
\font\fivesy=cmsy5	\skewchar\fivesy='60
\font\fivebf=cmbx5

\font\sevenrm=cmr7
\font\seveni=cmmi7	\skewchar\seveni='177
\font\sevensy=cmsy7	\skewchar\sevensy='60
\font\sevenbf=cmbx7

\font\eightrm=cmr8
\font\eightbf=cmbx8
\font\eightit=cmti8
\font\eighti=cmmi8			\skewchar\eighti='177
\font\eightmib=cmmib10 at 8pt	\skewchar\eightmib='177
\font\eightsy=cmsy8			\skewchar\eightsy='60
\font\eightsyb=cmbsy10 at 8pt	\skewchar\eightsyb='60
\font\eightsl=cmsl8
\font\eighttt=cmtt8			\hyphenchar\eighttt=-1
\font\eightcsc=cmcsc10 at 8pt
\font\eightsf=cmss8

\font\ninerm=cmr9
\font\ninebf=cmbx9
\font\nineit=cmti9
\font\ninei=cmmi9			\skewchar\ninei='177
\font\ninemib=cmmib10 at 9pt	\skewchar\ninemib='177
\font\ninesy=cmsy9			\skewchar\ninesy='60
\font\ninesyb=cmbsy10 at 9pt	\skewchar\ninesyb='60
\font\ninesl=cmsl9
\font\ninett=cmtt9			\hyphenchar\ninett=-1
\font\ninecsc=cmcsc10 at 9pt
\font\ninesf=cmss9

\font\tenrm=cmr10
\font\tenbf=cmbx10
\font\tenit=cmti10
\font\teni=cmmi10		\skewchar\teni='177
\font\tenmib=cmmib10	\skewchar\tenmib='177
\font\tensy=cmsy10		\skewchar\tensy='60
\font\tensyb=cmbsy10	\skewchar\tensyb='60
\font\tenex=cmex10
\font\tensl=cmsl10
\font\tentt=cmtt10		\hyphenchar\tentt=-1
\font\tencsc=cmcsc10
\font\tensf=cmss10

\font\elevenrm=cmr10 scaled \magstephalf
\font\elevenbf=cmbx10 scaled \magstephalf
\font\elevenit=cmti10 scaled \magstephalf
\font\eleveni=cmmi10 scaled \magstephalf	\skewchar\eleveni='177
\font\elevenmib=cmmib10 scaled \magstephalf	\skewchar\elevenmib='177
\font\elevensy=cmsy10 scaled \magstephalf	\skewchar\elevensy='60
\font\elevensyb=cmbsy10 scaled \magstephalf	\skewchar\elevensyb='60
\font\elevensl=cmsl10 scaled \magstephalf
\font\eleventt=cmtt10 scaled \magstephalf	\hyphenchar\eleventt=-1
\font\elevencsc=cmcsc10 scaled \magstephalf
\font\elevensf=cmss10 scaled \magstephalf

\font\fourteenrm=cmr10 scaled \magstep2
\font\fourteenbf=cmbx10 scaled \magstep2
\font\fourteenit=cmti10 scaled \magstep2
\font\fourteeni=cmmi10 scaled \magstep2		\skewchar\fourteeni='177
\font\fourteenmib=cmmib10 scaled \magstep2	\skewchar\fourteenmib='177
\font\fourteensy=cmsy10 scaled \magstep2	\skewchar\fourteensy='60
\font\fourteensyb=cmbsy10 scaled \magstep2	\skewchar\fourteensyb='60
\font\fourteensl=cmsl10 scaled \magstep2
\font\fourteentt=cmtt10 scaled \magstep2	\hyphenchar\fourteentt=-1
\font\fourteencsc=cmcsc10 scaled \magstep2
\font\fourteensf=cmss10 scaled \magstep2

\font\seventeenrm=cmr10 scaled \magstep3
\font\seventeenbf=cmbx10 scaled \magstep3
\font\seventeenit=cmti10 scaled \magstep3
\font\seventeeni=cmmi10 scaled \magstep3	\skewchar\seventeeni='177
\font\seventeenmib=cmmib10 scaled \magstep3	\skewchar\seventeenmib='177
\font\seventeensy=cmsy10 scaled \magstep3	\skewchar\seventeensy='60
\font\seventeensyb=cmbsy10 scaled \magstep3	\skewchar\seventeensyb='60
\font\seventeensl=cmsl10 scaled \magstep3
\font\seventeentt=cmtt10 scaled \magstep3	\hyphenchar\seventeentt=-1
\font\seventeencsc=cmcsc10 scaled \magstep3
\font\seventeensf=cmss10 scaled \magstep3

\def\@typeface{Computer Modern} 

\def\hexnumber@#1{\ifnum#1<10 \number#1\else
 \ifnum#1=10 A\else\ifnum#1=11 B\else\ifnum#1=12 C\else
 \ifnum#1=13 D\else\ifnum#1=14 E\else\ifnum#1=15 F\fi\fi\fi\fi\fi\fi\fi}

\def\mib{\hexnumber@\mibfam}
\def\syb{\hexnumber@\sybfam}

\def\makestrut{%
  \setbox\strutbox=\hbox{%
    \vrule height.7\baselineskip depth.3\baselineskip width 0pt}%
}

\def\bls#1{%
  \normalbaselineskip=#1%
  \normalbaselines%
  \makestrut%
}

%

\newfam\mibfam 
\newfam\sybfam 
\newfam\scfam  
\newfam\sffam  

\def\em{\ifdim\fontdimen1\font>0 \rm\else\it\fi}

\textfont3=\tenex
\scriptfont3=\tenex
\scriptscriptfont3=\tenex

\def\eightpoint{
  \def\rm{\fam0\eightrm}%
  \textfont0=\eightrm \scriptfont0=\sevenrm \scriptscriptfont0=\fiverm%
  \textfont1=\eighti  \scriptfont1=\seveni  \scriptscriptfont1=\fivei%
  \textfont2=\eightsy \scriptfont2=\sevensy \scriptscriptfont2=\fivesy%
  \textfont\itfam=\eightit\def\it{\fam\itfam\eightit}%
  \textfont\bffam=\eightbf%
    \scriptfont\bffam=\sevenbf%
      \scriptscriptfont\bffam=\fivebf%
  \def\bf{\fam\bffam\eightbf}%
  \textfont\slfam=\eightsl\def\sl{\fam\slfam\eightsl}%
  \textfont\ttfam=\eighttt\def\tt{\fam\ttfam\eighttt}%
  \textfont\scfam=\eightcsc\def\sc{\fam\scfam\eightcsc}%
  \textfont\sffam=\eightsf\def\sf{\fam\sffam\eightsf}%
  \textfont\mibfam=\eightmib%
  \textfont\sybfam=\eightsyb%
  \bls{10pt}%
}

\def\ninepoint{
  \def\rm{\fam0\ninerm}%
  \textfont0=\ninerm \scriptfont0=\sevenrm \scriptscriptfont0=\fiverm%
  \textfont1=\ninei  \scriptfont1=\seveni  \scriptscriptfont1=\fivei%
  \textfont2=\ninesy \scriptfont2=\sevensy \scriptscriptfont2=\fivesy%
  \textfont\itfam=\nineit\def\it{\fam\itfam\nineit}%
  \textfont\bffam=\ninebf%
    \scriptfont\bffam=\sevenbf%
      \scriptscriptfont\bffam=\fivebf%
  \def\bf{\fam\bffam\ninebf}%
  \textfont\slfam=\ninesl\def\sl{\fam\slfam\ninesl}%
  \textfont\ttfam=\ninett\def\tt{\fam\ttfam\ninett}%
  \textfont\scfam=\ninecsc\def\sc{\fam\scfam\ninecsc}%
  \textfont\sffam=\ninesf\def\sf{\fam\sffam\ninesf}%
  \textfont\mibfam=\ninemib%
  \textfont\sybfam=\ninesyb%
  \bls{12pt}%
}

\def\tenpoint{
  \def\rm{\fam0\tenrm}%
  \textfont0=\tenrm \scriptfont0=\sevenrm \scriptscriptfont0=\fiverm%
  \textfont1=\teni  \scriptfont1=\seveni  \scriptscriptfont1=\fivei%
  \textfont2=\tensy \scriptfont2=\sevensy \scriptscriptfont2=\fivesy%
  \textfont\itfam=\tenit\def\it{\fam\itfam\tenit}%
  \textfont\bffam=\tenbf%
    \scriptfont\bffam=\sevenbf%
      \scriptscriptfont\bffam=\fivebf%
  \def\bf{\fam\bffam\tenbf}%
  \textfont\slfam=\tensl\def\sl{\fam\slfam\tensl}%
  \textfont\ttfam=\tentt\def\tt{\fam\ttfam\tentt}%
  \textfont\scfam=\tencsc\def\sc{\fam\scfam\tencsc}%
  \textfont\sffam=\tensf\def\sf{\fam\sffam\tensf}%
  \textfont\mibfam=\tenmib%
  \textfont\sybfam=\tensyb%
  \bls{12pt}%
}

\def\elevenpoint{
  \def\rm{\fam0\elevenrm}%
  \textfont0=\elevenrm \scriptfont0=\eightrm \scriptscriptfont0=\fiverm%
  \textfont1=\eleveni  \scriptfont1=\eighti  \scriptscriptfont1=\fivei%
  \textfont2=\elevensy \scriptfont2=\eightsy \scriptscriptfont2=\fivesy%
  \textfont\itfam=\elevenit\def\it{\fam\itfam\elevenit}%
  \textfont\bffam=\elevenbf%
    \scriptfont\bffam=\eightbf%
      \scriptscriptfont\bffam=\fivebf%
  \def\bf{\fam\bffam\elevenbf}%
  \textfont\slfam=\elevensl\def\sl{\fam\slfam\elevensl}%
  \textfont\ttfam=\eleventt\def\tt{\fam\ttfam\eleventt}%
  \textfont\scfam=\elevencsc\def\sc{\fam\scfam\elevencsc}%
  \textfont\sffam=\elevensf\def\sf{\fam\sffam\elevensf}%
  \textfont\mibfam=\elevenmib%
  \textfont\sybfam=\elevensyb%
  \bls{13pt}%
}

\def\fourteenpoint{
  \def\rm{\fam0\fourteenrm}%
  \textfont0\fourteenrm  \scriptfont0\tenrm  \scriptscriptfont0\sevenrm%
  \textfont1\fourteeni   \scriptfont1\teni   \scriptscriptfont1\seveni%
  \textfont2\fourteensy  \scriptfont2\tensy  \scriptscriptfont2\sevensy%
  \textfont\itfam=\fourteenit\def\it{\fam\itfam\fourteenit}%
  \textfont\bffam=\fourteenbf%
    \scriptfont\bffam=\tenbf%
      \scriptscriptfont\bffam=\sevenbf%
  \def\bf{\fam\bffam\fourteenbf}%
  \textfont\slfam=\fourteensl\def\sl{\fam\slfam\fourteensl}%
  \textfont\ttfam=\fourteentt\def\tt{\fam\ttfam\fourteentt}%
  \textfont\scfam=\fourteencsc\def\sc{\fam\scfam\fourteencsc}%
  \textfont\sffam=\fourteensf\def\sf{\fam\sffam\fourteensf}%
  \textfont\mibfam=\fourteenmib%
  \textfont\sybfam=\fourteensyb%
  \bls{17pt}%
}

\def\seventeenpoint{
  \def\rm{\fam0\seventeenrm}%
  \textfont0\seventeenrm  \scriptfont0\elevenrm  \scriptscriptfont0\ninerm%
  \textfont1\seventeeni   \scriptfont1\eleveni   \scriptscriptfont1\ninei%
  \textfont2\seventeensy  \scriptfont2\elevensy  \scriptscriptfont2\ninesy%
  \textfont\itfam=\seventeenit\def\it{\fam\itfam\seventeenit}%
  \textfont\bffam=\seventeenbf%
    \scriptfont\bffam=\elevenbf%
      \scriptscriptfont\bffam=\ninebf%
  \def\bf{\fam\bffam\seventeenbf}%
  \textfont\slfam=\seventeensl\def\sl{\fam\slfam\seventeensl}%
  \textfont\ttfam=\seventeentt\def\tt{\fam\ttfam\seventeentt}%
  \textfont\scfam=\seventeencsc\def\sc{\fam\scfam\seventeencsc}%
  \textfont\sffam=\seventeensf\def\sf{\fam\sffam\seventeensf}%
  \textfont\mibfam=\seventeenmib%
  \textfont\sybfam=\seventeensyb%
  \bls{20pt}%
}

\lineskip=1pt      \normallineskip=\lineskip
\lineskiplimit=0pt \normallineskiplimit=\lineskiplimit




\def\Nulle{0}  
\def\Aue{1}    
\def\Afe{2}    
\def\Sue{4}    
\def\Hae{5}    
\def\Hbe{6}    
\def\Hce{7}    
\def\Hde{8}    
\def\Kwe{9}    
\def\Txe{10}   
\def\Lie{11}   
\def\Bbe{12}   


\newdimen\DimenA
\newbox\BoxA

\newcount\LastMac \LastMac=\Nulle
\newcount\HeaderNumber \HeaderNumber=0
\newcount\DefaultHeader \DefaultHeader=\HeaderNumber
\newskip\Indent

\newskip\half      \half=5.5pt plus 1.5pt minus 2.25pt
\newskip\one       \one=11pt plus 3pt minus 5.5pt
\newskip\onehalf   \onehalf=16.5pt plus 5.5pt minus 8.25pt
\newskip\two       \two=22pt plus 5.5pt minus 11pt

\def\Half{\vskip-\lastskip\vskip\half}
\def\One{\vskip-\lastskip\vskip\one}
\def\OneHalf{\vskip-\lastskip\vskip\onehalf}
\def\Two{\vskip-\lastskip\vskip\two}


\def\rTenPT{10pt plus \Feathering}

\def\TenPT{10pt plus \Feathering} 
\def\ElevenPT{11pt plus \Feathering}

\def\Raggedright{
 \rightskip=0pt plus \hsize
}

\def\Fullout{
\rightskip=0pt
}

\def\Hang#1#2{
 \hangindent=#1
 \hangafter=#2
}

\def\EveryMac{
 \Fullout
 \everypar{}
}



\def\title#1{
 \EveryMac
 \LastMac=\Nulle
 \global\HeaderNumber=0
 \global\DefaultHeader=1
 \vbox to 1pc{\vss}
 \seventeenpoint
 \Raggedright
 \noindent \bf #1
}

\def\author#1{
 \EveryMac
 \ifnum\LastMac=\Afe \OneHalf
  \else \Two
 \fi
 \LastMac=\Aue
 \fourteenpoint
 \Raggedright
 \noindent \rm #1\par
 \vskip 3pt\relax
}

\def\affiliation#1{
 \EveryMac
 \LastMac=\Afe
 \eightpoint\bls{\TenPT}
 \Raggedright
 \noindent \it #1\par
}

\def\abstract{%
 \EveryMac
 \Two
 \LastMac=\Sue
 \everypar{\Hang{11pc}{0}}
 \noindent\ninebf ABSTRACT\par
 \tenpoint\bls{\ElevenPT}
 \Fullout
 \noindent\rm
}

\def\keywords{
 \EveryMac
 \Half
 \LastMac=\Kwe
 \everypar{\Hang{11pc}{0}}
 \tenpoint\bls{\ElevenPT}
 \Fullout
 \noindent\hbox{\bf Key words:\ }
 \rm
}


\def\maketitle{%
  \Two%
  \EndOpening%
  \MakePage%
}


\def\pageoffset#1#2{\hoffset=#1\relax\voffset=#2\relax}


\def\Autonumber{
 \global\AutoNumbertrue  
}

\newif\ifAutoNumber \AutoNumberfalse
\newcount\Sec        
\newcount\SecSec
\newcount\SecSecSec

\Sec=0

\def\:{\let\@sptoken= } \:  
\def\:{\@xifnch} \expandafter\def\: {\futurelet\@tempc\@ifnch}

\def\@ifnextchar#1#2#3{%
  \let\@tempMACe #1%
  \def\@tempMACa{#2}%
  \def\@tempMACb{#3}%
  \futurelet \@tempMACc\@ifnch%
}

\def\@ifnch{%
\ifx \@tempMACc \@sptoken%
  \let\@tempMACd\@xifnch%
\else%
  \ifx \@tempMACc \@tempMACe%
    \let\@tempMACd\@tempMACa%
  \else%
    \let\@tempMACd\@tempMACb%
  \fi%
\fi%
\@tempMACd%
}

\def\@ifstar#1#2{\@ifnextchar *{\def\@tempMACa*{#1}\@tempMACa}{#2}}

\def\section{\@ifstar{\@ssection}{\@section}}

\def\@section#1{
 \EveryMac
 \Two
 \LastMac=\Hae
 \ninepoint\bls{\ElevenPT}
 \bf
 \Raggedright
 \ifAutoNumber
  \advance\Sec by 1
  \noindent\number\Sec\hskip 1pc \uppercase{#1}
  \SecSec=0
 \else
  \noindent \uppercase{#1}
 \fi
 \nobreak
}

\def\@ssection#1{
 \EveryMac
 \ifnum\LastMac=\Hae \Half
  \else \OneHalf
 \fi
 \LastMac=\Hae
 \tenpoint\bls{\ElevenPT}
 \bf
 \Raggedright
 \noindent\uppercase{#1}
}

\def\subsection#1{
 \EveryMac
 \ifnum\LastMac=\Hae \Half
  \else \OneHalf
 \fi
 \LastMac=\Hbe
 \tenpoint\bls{\ElevenPT}
 \bf
 \Raggedright
 \ifAutoNumber
  \advance\SecSec by 1
  \noindent\number\Sec.\number\SecSec
  \hskip 1pc #1
  \SecSecSec=0
 \else
  \noindent #1
 \fi
 \nobreak
}

\def\subsubsection#1{
 \EveryMac
 \ifnum\LastMac=\Hbe \Half
  \else \OneHalf
 \fi
 \LastMac=\Hce
 \ninepoint\bls{\ElevenPT}
 \it
 \Raggedright
 \ifAutoNumber
  \advance\SecSecSec by 1
  \noindent\number\Sec.\number\SecSec.\number\SecSecSec
  \hskip 1pc #1
 \else
  \noindent #1
 \fi
 \nobreak
}

\def\paragraph#1{
 \EveryMac
 \One
 \LastMac=\Hde
 \ninepoint\bls{\ElevenPT}
 \noindent \it #1
 \rm
}


\def\tx{
 \EveryMac
 \ifnum\LastMac=\Lie \Half\fi
 \ifnum\LastMac=\Hae \nobreak\Half\fi
 \ifnum\LastMac=\Hbe \nobreak\Half\fi
 \ifnum\LastMac=\Hce \nobreak\Half\fi
 \ifnum\LastMac=\Lie \else \noindent\fi
 \LastMac=\Txe
 \ninepoint\bls{\ElevenPT}
 \rm
}


\def\item{
 \par
 \EveryMac
 \ifnum\LastMac=\Lie
  \else \Half
 \fi
 \LastMac=\Lie
 \ninepoint\bls{\ElevenPT}
 \rm
}


\def\bibitem{
 \par
 \EveryMac
 \ifnum\LastMac=\Bbe
  \else \Half
 \fi
 \LastMac=\Bbe
 \Hang{1.5em}{1}
 \eightpoint\bls{\TenPT}
 \Raggedright
 \noindent \rm
}


\newtoks\CatchLine

\def\@journal{Mon.\ Not.\ R.\ Astron.\ Soc.\ }  
\def\@pubyear{1993}        
\def\@pagerange{000--000}  
\def\@volume{000}          
\def\@microfiche{}         %

\def\pubyear#1{\gdef\@pubyear{#1}\@makecatchline}
\def\pagerange#1{\gdef\@pagerange{#1}\@makecatchline}
\def\volume#1{\gdef\@volume{#1}\@makecatchline}
\def\microfiche#1{\gdef\@microfiche{and Microfiche\ #1}\@makecatchline}

\def\@makecatchline{%
  \global\CatchLine{%
    {\rm \@journal {\bf \@volume},\ \@pagerange\ (\@pubyear)\ \@microfiche}}%
}

\@makecatchline 

\newtoks\LeftHeader
\def\shortauthor#1{
 \global\LeftHeader{#1}
}

\newtoks\RightHeader
\def\shorttitle#1{
 \global\RightHeader{#1}
}

\def\PageHead{
 \EveryMac
 \ifnum\HeaderNumber=1 \Pagehead
  \else \Catchline
 \fi
}

\def\Catchline{%
 \vbox to 0pt{\vskip-22.5pt
  \hbox to \PageWidth{\vbox to8.5pt{}\noindent
  \eightpoint\the\CatchLine\hfill}\vss}
 \nointerlineskip
}

\def\Pagehead{%
 \ifodd\pageno
   \vbox to 0pt{\vskip-22.5pt
   \hbox to \PageWidth{\vbox to8.5pt{}\elevenpoint\it\noindent
    \hfill\the\RightHeader\hskip1.5em\rm\folio}\vss}
 \else
   \vbox to 0pt{\vskip-22.5pt
   \hbox to \PageWidth{\vbox to8.5pt{}\elevenpoint\rm\noindent
   \folio\hskip1.5em\it\the\LeftHeader\hfill}\vss}
 \fi
 \nointerlineskip
}

\def\PageFoot{} 

\def\authorcomment#1{%
  \gdef\PageFoot{%
    \nointerlineskip%
    \vbox to 22pt{\vfil%
      \hbox to \PageWidth{\elevenpoint\rm\noindent \hfil #1 \hfil}}%
  }%
}

\everydisplay{\displaysetup}

\newif\ifeqno
\newif\ifleqno

\def\displaysetup#1$${%
 \displaytest#1\eqno\eqno\displaytest
}

\def\displaytest#1\eqno#2\eqno#3\displaytest{%
 \if!#3!\ldisplaytest#1\leqno\leqno\ldisplaytest
 \else\eqnotrue\leqnofalse\def\eqn{#2}\def\eq{#1}\fi
 \generaldisplay$$}

\def\ldisplaytest#1\leqno#2\leqno#3\ldisplaytest{%
 \def\eq{#1}%
 \if!#3!\eqnofalse\else\eqnotrue\leqnotrue
  \def\eqn{#2}\fi}

\def\generaldisplay{%
\ifeqno \ifleqno 
   \hbox to \hsize{\noindent
     $\displaystyle\eq$\hfil$\displaystyle\eqn$}
  \else
    \hbox to \hsize{\noindent
     $\displaystyle\eq$\hfil$\displaystyle\eqn$}
  \fi
 \else
 \hbox to \hsize{\vbox{\noindent
  $\displaystyle\eq$\hfil}}
 \fi
}

\def\@notice{%
  \par\Two%
  \bls{12pt}%
  \noindent\tenrm This paper has been produced using the Blackwell
                  Scientific Publications \TeX\ macros.%
}

\outer\def\bye{\@notice\par\vfill\supereject\end}

\everyjob{%
  \Warn{Monthly notices of the RAS journal style (\@typeface)\space
        v\@version,\space \@verdate.}\Warn{}%
}




\newif\if@debug \@debugfalse  

\def\Print#1{\if@debug\immediate\write16{#1}\else \fi}
\def\Warn#1{\immediate\write16{#1}}
\def\wlog#1{}

\newcount\Iteration 

\newif\ifFigureBoxes  
\FigureBoxestrue

\def\Single{0} \def\Double{1}                 
\def\Figure{0} \def\Table{1}                  

\def\InStack{0}  
\def\InZoneA{1}
\def\InZoneB{2}
\def\InZoneC{3}

\newcount\TEMPCOUNT 
\newdimen\TEMPDIMEN 
\newbox\TEMPBOX     
\newbox\VOIDBOX     

\newcount\LengthOfStack 
\newcount\MaxItems      
\newcount\StackPointer
\newcount\Point         
\newcount\NextFigure    
\newcount\NextTable     
\newcount\NextItem      

\newcount\StatusStack   
\newcount\NumStack      
\newcount\TypeStack     
\newcount\SpanStack     
\newcount\BoxStack      

\newcount\ItemSTATUS    
\newcount\ItemNUMBER    
\newcount\ItemTYPE      
\newcount\ItemSPAN      
\newbox\ItemBOX         
\newdimen\ItemSIZE      

\newdimen\PageHeight    
\newdimen\TextLeading   
\newdimen\Feathering    
\newcount\LinesPerPage  
\newdimen\ColumnWidth   
\newdimen\ColumnGap     
\newdimen\PageWidth     
\newdimen\BodgeHeight   
\newcount\Leading       

\newdimen\ZoneBSize  
\newdimen\TextSize   
\newbox\ZoneABOX     
\newbox\ZoneBBOX     
\newbox\ZoneCBOX     

\newif\ifFirstSingleItem
\newif\ifFirstZoneA
\newif\ifMakePageInComplete
\newif\ifMoreFigures \MoreFiguresfalse 
\newif\ifMoreTables  \MoreTablesfalse  

\newif\ifFigInZoneB 
\newif\ifFigInZoneC 
\newif\ifTabInZoneB 
\newif\ifTabInZoneC

\newif\ifZoneAFullPage

\newbox\MidBOX    
\newbox\LeftBOX
\newbox\RightBOX
\newbox\PageBOX   

\newif\ifLeftCOL  
\LeftCOLtrue

\newdimen\ZoneBAdjust

\newcount\ItemFits
\def\Yes{1}
\def\No{2}




\MaxItems=15
\NextFigure=0        
\NextTable=1

\BodgeHeight=6pt
\TextLeading=11pt    
\Leading=11
\Feathering=0pt      
\LinesPerPage=61     
\topskip=\TextLeading
\ColumnWidth=20pc    
\ColumnGap=2pc       

\def\ItemSep{\vskip \TextLeading plus \TextLeading minus 4pt}

\FigureBoxesfalse 

\parskip=0pt
\parindent=18pt
\widowpenalty=0
\clubpenalty=10000
\tolerance=1500
\hbadness=1500
\abovedisplayskip=6pt plus 2pt minus 2pt
\belowdisplayskip=6pt plus 2pt minus 2pt
\abovedisplayshortskip=6pt plus 2pt minus 2pt
\belowdisplayshortskip=6pt plus 2pt minus 2pt

\PageHeight=\TextLeading 
\multiply\PageHeight by \LinesPerPage
\advance\PageHeight by \topskip

\PageWidth=2\ColumnWidth
\advance\PageWidth by \ColumnGap




\newcount\DUMMY \StatusStack=\allocationnumber
\newcount\DUMMY \newcount\DUMMY \newcount\DUMMY 
\newcount\DUMMY \newcount\DUMMY \newcount\DUMMY 
\newcount\DUMMY \newcount\DUMMY \newcount\DUMMY
\newcount\DUMMY \newcount\DUMMY \newcount\DUMMY 
\newcount\DUMMY \newcount\DUMMY \newcount\DUMMY

\newcount\DUMMY \NumStack=\allocationnumber
\newcount\DUMMY \newcount\DUMMY \newcount\DUMMY 
\newcount\DUMMY \newcount\DUMMY \newcount\DUMMY 
\newcount\DUMMY \newcount\DUMMY \newcount\DUMMY 
\newcount\DUMMY \newcount\DUMMY \newcount\DUMMY 
\newcount\DUMMY \newcount\DUMMY \newcount\DUMMY

\newcount\DUMMY \TypeStack=\allocationnumber
\newcount\DUMMY \newcount\DUMMY \newcount\DUMMY 
\newcount\DUMMY \newcount\DUMMY \newcount\DUMMY 
\newcount\DUMMY \newcount\DUMMY \newcount\DUMMY 
\newcount\DUMMY \newcount\DUMMY \newcount\DUMMY 
\newcount\DUMMY \newcount\DUMMY \newcount\DUMMY

\newcount\DUMMY \SpanStack=\allocationnumber
\newcount\DUMMY \newcount\DUMMY \newcount\DUMMY 
\newcount\DUMMY \newcount\DUMMY \newcount\DUMMY 
\newcount\DUMMY \newcount\DUMMY \newcount\DUMMY 
\newcount\DUMMY \newcount\DUMMY \newcount\DUMMY 
\newcount\DUMMY \newcount\DUMMY \newcount\DUMMY

\newbox\DUMMY   \BoxStack=\allocationnumber
\newbox\DUMMY   \newbox\DUMMY \newbox\DUMMY 
\newbox\DUMMY   \newbox\DUMMY \newbox\DUMMY 
\newbox\DUMMY   \newbox\DUMMY \newbox\DUMMY 
\newbox\DUMMY   \newbox\DUMMY \newbox\DUMMY 
\newbox\DUMMY   \newbox\DUMMY \newbox\DUMMY

\def\wlog{\immediate\write-1}


\def\GetItemAll#1{%
 \GetItemSTATUS{#1}
 \GetItemNUMBER{#1}
 \GetItemTYPE{#1}
 \GetItemSPAN{#1}
 \GetItemBOX{#1}
}

\def\GetItemSTATUS#1{%
 \Point=\StatusStack
 \advance\Point by #1
 \global\ItemSTATUS=\count\Point
}

\def\GetItemNUMBER#1{%
 \Point=\NumStack
 \advance\Point by #1
 \global\ItemNUMBER=\count\Point
}

\def\GetItemTYPE#1{%
 \Point=\TypeStack
 \advance\Point by #1
 \global\ItemTYPE=\count\Point
}

\def\GetItemSPAN#1{%
 \Point\SpanStack
 \advance\Point by #1
 \global\ItemSPAN=\count\Point
}

\def\GetItemBOX#1{%
 \Point=\BoxStack
 \advance\Point by #1
 \global\setbox\ItemBOX=\vbox{\copy\Point}
 \global\ItemSIZE=\ht\ItemBOX
 \global\advance\ItemSIZE by \dp\ItemBOX
 \TEMPCOUNT=\ItemSIZE
 \divide\TEMPCOUNT by \Leading
 \divide\TEMPCOUNT by 65536
 \advance\TEMPCOUNT by 1
 \ItemSIZE=\TEMPCOUNT pt
 \global\multiply\ItemSIZE by \Leading
}


\def\JoinStack{%
 \ifnum\LengthOfStack=\MaxItems 
  \Warn{WARNING: Stack is full...some items will be lost!}
 \else
  \Point=\StatusStack
  \advance\Point by \LengthOfStack
  \global\count\Point=\ItemSTATUS
  \Point=\NumStack
  \advance\Point by \LengthOfStack
  \global\count\Point=\ItemNUMBER
  \Point=\TypeStack
  \advance\Point by \LengthOfStack
  \global\count\Point=\ItemTYPE
  \Point\SpanStack
  \advance\Point by \LengthOfStack
  \global\count\Point=\ItemSPAN
  \Point=\BoxStack
  \advance\Point by \LengthOfStack
  \global\setbox\Point=\vbox{\copy\ItemBOX}
  \global\advance\LengthOfStack by 1
  \ifnum\ItemTYPE=\Figure 
   \global\MoreFigurestrue
  \else
   \global\MoreTablestrue
  \fi
 \fi
}


\def\LeaveStack#1{%
 {\Iteration=#1
 \loop
 \ifnum\Iteration<\LengthOfStack
  \advance\Iteration by 1
  \GetItemSTATUS{\Iteration}
   \advance\Point by -1
   \global\count\Point=\ItemSTATUS
  \GetItemNUMBER{\Iteration}
   \advance\Point by -1
   \global\count\Point=\ItemNUMBER
  \GetItemTYPE{\Iteration}
   \advance\Point by -1
   \global\count\Point=\ItemTYPE
  \GetItemSPAN{\Iteration}
   \advance\Point by -1
   \global\count\Point=\ItemSPAN
  \GetItemBOX{\Iteration}
   \advance\Point by -1
   \global\setbox\Point=\vbox{\copy\ItemBOX}
 \repeat}
 \global\advance\LengthOfStack by -1
}


\newif\ifStackNotClean

\def\CleanStack{%
 \StackNotCleantrue
 {\Iteration=0
  \loop
   \ifStackNotClean
    \GetItemSTATUS{\Iteration}
    \ifnum\ItemSTATUS=\InStack
     \advance\Iteration by 1
     \else
      \LeaveStack{\Iteration}
    \fi
   \ifnum\LengthOfStack<\Iteration
    \StackNotCleanfalse
   \fi
 \repeat}
}


\def\FindItem#1#2{%
 \global\StackPointer=-1 
 {\Iteration=0
  \loop
  \ifnum\Iteration<\LengthOfStack
   \GetItemSTATUS{\Iteration}
   \ifnum\ItemSTATUS=\InStack
    \GetItemTYPE{\Iteration}
    \ifnum\ItemTYPE=#1
     \GetItemNUMBER{\Iteration}
     \ifnum\ItemNUMBER=#2
      \global\StackPointer=\Iteration
      \Iteration=\LengthOfStack 
     \fi
    \fi
   \fi
  \advance\Iteration by 1
 \repeat}
}


\def\FindNext{%
 \global\StackPointer=-1 
 {\Iteration=0
  \loop
  \ifnum\Iteration<\LengthOfStack
   \GetItemSTATUS{\Iteration}
   \ifnum\ItemSTATUS=\InStack
    \GetItemTYPE{\Iteration}
   \ifnum\ItemTYPE=\Figure
    \ifMoreFigures
      \global\NextItem=\Figure
      \global\StackPointer=\Iteration
      \Iteration=\LengthOfStack 
    \fi
   \fi
   \ifnum\ItemTYPE=\Table
    \ifMoreTables
      \global\NextItem=\Table
      \global\StackPointer=\Iteration
      \Iteration=\LengthOfStack 
    \fi
   \fi
  \fi
  \advance\Iteration by 1
 \repeat}
}


\def\ChangeStatus#1#2{%
 \Point=\StatusStack
 \advance\Point by #1
 \global\count\Point=#2
}



\def\Zone{\InZoneA}

\ZoneBAdjust=0pt

\def\MakePage{
 \global\ZoneBSize=\PageHeight
 \global\TextSize=\ZoneBSize
 \global\ZoneAFullPagefalse
 \global\topskip=\TextLeading
 \MakePageInCompletetrue
 \MoreFigurestrue
 \MoreTablestrue
 \FigInZoneBfalse
 \FigInZoneCfalse
 \TabInZoneBfalse
 \TabInZoneCfalse
 \global\FirstSingleItemtrue
 \global\FirstZoneAtrue
 \global\setbox\ZoneABOX=\box\VOIDBOX
 \global\setbox\ZoneBBOX=\box\VOIDBOX
 \global\setbox\ZoneCBOX=\box\VOIDBOX
 \loop
  \ifMakePageInComplete
 \FindNext
 \ifnum\StackPointer=-1
  \NextItem=-1
  \MoreFiguresfalse
  \MoreTablesfalse
 \fi
 \ifnum\NextItem=\Figure
   \FindItem{\Figure}{\NextFigure}
   \ifnum\StackPointer=-1 \global\MoreFiguresfalse
   \else
    \GetItemSPAN{\StackPointer}
    \ifnum\ItemSPAN=\Single \def\Zone{\InZoneB}\relax
     \ifFigInZoneC \global\MoreFiguresfalse\fi
    \else
     \def\Zone{\InZoneA}
     \ifFigInZoneB \def\Zone{\InZoneC}\fi
    \fi
   \fi
   \ifMoreFigures\Print{}\FigureItems\fi
 \fi
\ifnum\NextItem=\Table
   \FindItem{\Table}{\NextTable}
   \ifnum\StackPointer=-1 \global\MoreTablesfalse
   \else
    \GetItemSPAN{\StackPointer}
    \ifnum\ItemSPAN=\Single\relax
     \ifTabInZoneC \global\MoreTablesfalse\fi
    \else
     \def\Zone{\InZoneA}
     \ifTabInZoneB \def\Zone{\InZoneC}\fi
    \fi
   \fi
   \ifMoreTables\Print{}\TableItems\fi
 \fi
   \MakePageInCompletefalse 
   \ifMoreFigures\MakePageInCompletetrue\fi
   \ifMoreTables\MakePageInCompletetrue\fi
 \repeat
 \ifZoneAFullPage
  \global\TextSize=0pt
  \global\ZoneBSize=0pt
  \global\vsize=0pt\relax
  \global\topskip=0pt\relax
  \vbox to 0pt{\vss}
  \eject
 \else
 \global\advance\ZoneBSize by -\ZoneBAdjust
 \global\vsize=\ZoneBSize
 \global\hsize=\ColumnWidth
 \global\ZoneBAdjust=0pt
 \ifdim\TextSize<23pt
 \Warn{}
 \Warn{* Making column fall short: TextSize=\the\TextSize *}
 \vskip-\lastskip\eject\fi
 \fi
}

\def\MakeRightCol{
 \global\TextSize=\ZoneBSize
 \MakePageInCompletetrue
 \MoreFigurestrue
 \MoreTablestrue
 \global\FirstSingleItemtrue
 \global\setbox\ZoneBBOX=\box\VOIDBOX
 \def\Zone{\InZoneB}
 \loop
  \ifMakePageInComplete
 \FindNext
 \ifnum\StackPointer=-1
  \NextItem=-1
  \MoreFiguresfalse
  \MoreTablesfalse
 \fi
 \ifnum\NextItem=\Figure
   \FindItem{\Figure}{\NextFigure}
   \ifnum\StackPointer=-1 \MoreFiguresfalse
   \else
    \GetItemSPAN{\StackPointer}
    \ifnum\ItemSPAN=\Double\relax
     \MoreFiguresfalse\fi
   \fi
   \ifMoreFigures\Print{}\FigureItems\fi
 \fi
 \ifnum\NextItem=\Table
   \FindItem{\Table}{\NextTable}
   \ifnum\StackPointer=-1 \MoreTablesfalse
   \else
    \GetItemSPAN{\StackPointer}
    \ifnum\ItemSPAN=\Double\relax
     \MoreTablesfalse\fi
   \fi
   \ifMoreTables\Print{}\TableItems\fi
 \fi
   \MakePageInCompletefalse 
   \ifMoreFigures\MakePageInCompletetrue\fi
   \ifMoreTables\MakePageInCompletetrue\fi
 \repeat
 \ifZoneAFullPage
  \global\TextSize=0pt
  \global\ZoneBSize=0pt
  \global\vsize=0pt\relax
  \global\topskip=0pt\relax
  \vbox to 0pt{\vss}
  \eject
 \else
 \global\vsize=\ZoneBSize
 \global\hsize=\ColumnWidth
 \ifdim\TextSize<23pt
 \Warn{}
 \Warn{* Making column fall short: TextSize=\the\TextSize *}
 \vskip-\lastskip\eject\fi
\fi
}

\def\FigureItems{
 \Print{Considering...}
 \ShowItem{\StackPointer}
 \GetItemBOX{\StackPointer} 
 \GetItemSPAN{\StackPointer}
  \CheckFitInZone 
  \ifnum\ItemFits=\Yes
   \ifnum\ItemSPAN=\Single
     \ChangeStatus{\StackPointer}{\InZoneB} 
     \global\FigInZoneBtrue
     \ifFirstSingleItem
      \hbox{}\vskip-\BodgeHeight
     \global\advance\ItemSIZE by \TextLeading
     \fi
     \unvbox\ItemBOX\ItemSep
     \global\FirstSingleItemfalse
     \global\advance\TextSize by -\ItemSIZE
     \global\advance\TextSize by -\TextLeading
   \else
    \ifFirstZoneA
     \global\advance\ItemSIZE by \TextLeading
     \global\FirstZoneAfalse\fi
    \global\advance\TextSize by -\ItemSIZE
    \global\advance\TextSize by -\TextLeading
    \global\advance\ZoneBSize by -\ItemSIZE
    \global\advance\ZoneBSize by -\TextLeading
    \ifFigInZoneB\relax
     \else
     \ifdim\TextSize<3\TextLeading
     \global\ZoneAFullPagetrue
     \fi
    \fi
    \ChangeStatus{\StackPointer}{\Zone}
    \ifnum\Zone=\InZoneC \global\FigInZoneCtrue\fi
  \fi
   \Print{TextSize=\the\TextSize}
   \Print{ZoneBSize=\the\ZoneBSize}
  \global\advance\NextFigure by 1
   \Print{This figure has been placed.}
  \else
   \Print{No space available for this figure...holding over.}
   \Print{}
   \global\MoreFiguresfalse
  \fi
}

\def\TableItems{
 \Print{Considering...}
 \ShowItem{\StackPointer}
 \GetItemBOX{\StackPointer} 
 \GetItemSPAN{\StackPointer}
  \CheckFitInZone 
  \ifnum\ItemFits=\Yes
   \ifnum\ItemSPAN=\Single
    \ChangeStatus{\StackPointer}{\InZoneB}
     \global\TabInZoneBtrue
     \ifFirstSingleItem
      \hbox{}\vskip-\BodgeHeight
     \global\advance\ItemSIZE by \TextLeading
     \fi
     \unvbox\ItemBOX\ItemSep
     \global\FirstSingleItemfalse
     \global\advance\TextSize by -\ItemSIZE
     \global\advance\TextSize by -\TextLeading
   \else
    \ifFirstZoneA
    \global\advance\ItemSIZE by \TextLeading
    \global\FirstZoneAfalse\fi
    \global\advance\TextSize by -\ItemSIZE
    \global\advance\TextSize by -\TextLeading
    \global\advance\ZoneBSize by -\ItemSIZE
    \global\advance\ZoneBSize by -\TextLeading
    \ifFigInZoneB\relax
     \else
     \ifdim\TextSize<3\TextLeading
     \global\ZoneAFullPagetrue
     \fi
    \fi
    \ChangeStatus{\StackPointer}{\Zone}
    \ifnum\Zone=\InZoneC \global\TabInZoneCtrue\fi
   \fi
  \global\advance\NextTable by 1
   \Print{This table has been placed.}
  \else
  \Print{No space available for this table...holding over.}
   \Print{}
   \global\MoreTablesfalse
  \fi
}


\def\CheckFitInZone{%
{\advance\TextSize by -\ItemSIZE
 \advance\TextSize by -\TextLeading
 \ifFirstSingleItem
  \advance\TextSize by \TextLeading
 \fi
 \ifnum\Zone=\InZoneA\relax
  \else \advance\TextSize by -\ZoneBAdjust
 \fi
 \ifdim\TextSize<3\TextLeading \global\ItemFits=\No
 \else \global\ItemFits=\Yes\fi}
}

\def\BF#1#2{
 \ItemSTATUS=\InStack
 \ItemNUMBER=#1
 \ItemTYPE=\Figure
 \if#2S \ItemSPAN=\Single
  \else \ItemSPAN=\Double
 \fi
 \setbox\ItemBOX=\vbox{}
}

\def\BT#1#2{
 \ItemSTATUS=\InStack
 \ItemNUMBER=#1
 \ItemTYPE=\Table
 \if#2S \ItemSPAN=\Single
  \else \ItemSPAN=\Double
 \fi
 \setbox\ItemBOX=\vbox{}
}

\def\BeginOpening{%
 \hsize=\PageWidth
 \global\setbox\ItemBOX=\vbox\bgroup
}

\let\begintopmatter=\BeginOpening  

\def\EndOpening{%
 \egroup
 \ItemNUMBER=0
 \ItemTYPE=\Figure
 \ItemSPAN=\Double
 \ItemSTATUS=\InStack
 \JoinStack
}


\newbox\tmpbox

\def\FC#1#2#3#4{%
  \ItemSTATUS=\InStack
  \ItemNUMBER=#1
  \ItemTYPE=\Figure
  \if#2S
    \ItemSPAN=\Single \TEMPDIMEN=\ColumnWidth
  \else
    \ItemSPAN=\Double \TEMPDIMEN=\PageWidth
  \fi
  {\hsize=\TEMPDIMEN
   \global\setbox\ItemBOX=\vbox{%
     \ifFigureBoxes
       \B{\TEMPDIMEN}{#3}
     \else
       \vbox to #3{\vfil}%
     \fi%
     \eightpoint\rm\bls{\rTenPT}%
     \vskip 5.5pt plus 6pt%
     \setbox\tmpbox=\vbox{#4\par}%
     \ifdim\ht\tmpbox>10pt 
       \noindent #4\par%
     \else
       \hbox to \hsize{\hfil #4\hfil}%
     \fi%
   }%
  }%
  \JoinStack%
  \Print{Processing source for figure {\the\ItemNUMBER}}%
}

\let\figure=\FC  

\def\TH#1#2#3#4{%
 \ItemSTATUS=\InStack
 \ItemNUMBER=#1
 \ItemTYPE=\Table
 \if#2S \ItemSPAN=\Single \TEMPDIMEN=\ColumnWidth
  \else \ItemSPAN=\Double \TEMPDIMEN=\PageWidth
 \fi
{\hsize=\TEMPDIMEN
\eightpoint\bls{\rTenPT}\rm
\global\setbox\ItemBOX=\vbox{\noindent#3\vskip 5.5pt plus5.5pt\noindent#4}}
 \JoinStack
 \Print{Processing source for table {\the\ItemNUMBER}}
}

\let\table=\TH  

\def\UnloadZoneA{%
\FirstZoneAtrue
 \Iteration=0
  \loop
   \ifnum\Iteration<\LengthOfStack
    \GetItemSTATUS{\Iteration}
    \ifnum\ItemSTATUS=\InZoneA
     \GetItemBOX{\Iteration}
     \ifFirstZoneA \vbox to \BodgeHeight{\vfil}%
     \FirstZoneAfalse\fi
     \unvbox\ItemBOX\ItemSep
     \LeaveStack{\Iteration}
     \else
     \advance\Iteration by 1
   \fi
 \repeat
}

\def\UnloadZoneC{%
\Iteration=0
  \loop
   \ifnum\Iteration<\LengthOfStack
    \GetItemSTATUS{\Iteration}
    \ifnum\ItemSTATUS=\InZoneC
     \GetItemBOX{\Iteration}
     \ItemSep\unvbox\ItemBOX
     \LeaveStack{\Iteration}
     \else
     \advance\Iteration by 1
   \fi
 \repeat
}


\def\ShowItem#1{
  {\GetItemAll{#1}
  \Print{\the#1:
  {TYPE=\ifnum\ItemTYPE=\Figure Figure\else Table\fi}
  {NUMBER=\the\ItemNUMBER}
  {SPAN=\ifnum\ItemSPAN=\Single Single\else Double\fi}
  {SIZE=\the\ItemSIZE}}}
}

\def\ShowStack{%
 \Print{}
 \Print{LengthOfStack = \the\LengthOfStack}
 \ifnum\LengthOfStack=0 \Print{Stack is empty}\fi
 \Iteration=0
 \loop
 \ifnum\Iteration<\LengthOfStack
  \ShowItem{\Iteration}
  \advance\Iteration by 1
 \repeat
}

\def\B#1#2{%
\hbox{\vrule\kern-0.4pt\vbox to #2{%
\hrule width #1\vfill\hrule}\kern-0.4pt\vrule}
}

\def\Ref#1{\begingroup\global\setbox\TEMPBOX=\vbox{\hsize=2in\noindent#1}\endgroup
\ht1=0pt\dp1=0pt\wd1=0pt\vadjust{\vtop to 0pt{\advance
\hsize0.5pc\kern-10pt\moveright\hsize\box\TEMPBOX\vss}}}

\def\MarkRef#1{\leavevmode\thinspace\hbox{\vrule\vtop
{\vbox{\hrule\kern1pt\hbox{\vphantom{\rm/}\thinspace{\rm#1}%
\thinspace}}\kern1pt\hrule}\vrule}\thinspace}%


\output{%
 \ifLeftCOL
  \global\setbox\LeftBOX=\vbox to \ZoneBSize{\box255\unvbox\ZoneBBOX}
  \global\LeftCOLfalse
  \MakeRightCol
 \else
  \setbox\RightBOX=\vbox to \ZoneBSize{\box255\unvbox\ZoneBBOX}
  \setbox\MidBOX=\hbox{\box\LeftBOX\hskip\ColumnGap\box\RightBOX}
  \setbox\PageBOX=\vbox to \PageHeight{%
  \UnloadZoneA\box\MidBOX\UnloadZoneC}
  \shipout\vbox{\PageHead\box\PageBOX\PageFoot}
  \global\advance\pageno by 1
  \global\HeaderNumber=\DefaultHeader
  \global\LeftCOLtrue
  \CleanStack
  \MakePage
 \fi
}


\catcode `\@=12 
\def\210{{\rm 2--10\,keV}}


\pageoffset{-2pc}{0pc}
\Autonumber
\begintopmatter
\title{{\it ASCA} observations of Deep ROSAT fields - III.  The 
Discovery of an Obscured Type 2 AGN at $z=0.67$}
\author{B.J.Boyle$^1$, O.Almaini$^2$, I.Georgantopoulos$^3$, 
A.J.Blair$^4$, G.C.Stewart$^4$, R.E.Griffiths$^5$, T.Shanks$^6$, 
K.F.Gunn$^6$}
\affiliation{$^1$ Anglo-Australian Observatory, PO Box 296, Epping, 
NSW 2121, Australia}
\affiliation{$^2$ Institute for Astronomy, University of Edinburgh,
Blackford Hill, Edinburgh, EH9 3HJ}
\affiliation{$^3$ National Observatory of Athens, Lofos Koufou,
Palaia Penteli, 15236, Athens, Greece}
\affiliation{$^4$ Department of Physics \& Astronomy, The University 
of Leicester, Leicester LE1 7RH}
\affiliation{$^5$ Department of Physics, Carnegie Mellon University, 
Wean Hall, 5000 Forbes Ave., Pittsburgh, PA 15213, USA}
\affiliation{$^6$ Physics Department, University of Durham, South Road,
Durham DH1 3LE}

\shortauthor{B.J.Boyle et al.}
\shorttitle{{\it ASCA} - III: A Type 2 AGN at $z=0.67$}

\abstract 

We report on the discovery of a narrow-emission-line object at
$z=0.672$ detected in a deep {\it ASCA} survey.  The object,
AXJ0341.4--4453, has a flux in the 2--10$\,$keV band of $1.1\pm0.27
\times 10^{-13}\,$erg$\,$s$^{-1}$cm$^{-2}$, corresponding to a
luminosity of $1.8 \times 10^{44}\,$erg$\,$s$^{-1}$ ($q_0=0.5$, $H_0 =
50\,$km$\,$s$^{-1}$Mpc$^{-1}$). It is also marginally detected in the
{\it ROSAT} 0.5--2$\,$keV band with a flux $5.8 \times
10^{-15}\,$erg$\,$s$^{-1}$cm$^{-2}$. Both the {\it ASCA} data alone
and the combined {\it ROSAT/ASCA} data show a very hard X-ray
spectrum, consistent with either a flat power law ($\alpha < 0.1$) or
photoelectric absorption with a column of $n_{\rm H} > 4 \times
10^{22}\,$cm$^{-2}$ ($\alpha = 1$).  The optical spectrum shows the
high-ionisation, narrow emission lines typical of a Seyfert 2 galaxy.
We suggest that this object may be typical of the hard sources
required to explain the remainder of the X-ray background at hard
energies.

\keywords X-rays: general -- galaxies: active -- quasars: general
\maketitle
 
\section{Introduction}\tx

The high-redshift analogues to Seyfert 2 galaxies, the so-called Type
2 or obscured active galactic nuclei (AGN), have been increasingly
postulated as a major contributor to the X-ray background (XRB) in
recent years (see e.g. Comastri et al.\ 1995). Interest in these
objects has centred on their X-ray spectra, which are likely to be
significantly harder than those observed in normal QSOs or Type 1
AGN. Ordinary QSOs are known to comprise the bulk of the soft
($0.5-2\,$keV) source population identified to date (Shanks et al.\
1991, Schmidt et al.\ 1998) but their spectra are too soft to explain
the remainder of the XRB at harder energies.  Even relatively modest
amounts of gas and dust ($n_{\rm H} > 10^{21}\,$cm$^{-2}$) in the
central regions would readily absorb soft X-ray photons ($<2\,$keV)
and significantly harden the X-ray spectrum. A population of objects
with X-ray spectra of this type can readily reproduce the spectrum of
the X-ray background (Madau et al.\ 1994, Comastri et al.\ 1995).  The
discovery of at least one obscured QSO at high redshift ($z=2.35$)
lends some weight to this hypothesis (Almaini et al.\ 1995), 
although the unambiguous identification of large numbers
of such objects remains elusive.

A number of groups (Boyle et al.\ 1995a, Griffiths et al.\ 1996) have
detected increasing numbers of low redshift ($z<0.5$) galaxies with
narrow emission lines (NELGs) as the optical counterparts to faint
{\it ROSAT} sources.  The precise nature of these objects is unclear
(Boyle et al.\ 1995b); optical classification based on standard
emission-line-ratio diagnostics reveals them to be borderline
starburst/Seyfert 2 galaxies.  A hybrid explanation consisting of both
phenomena is a strong possibility (Boyle et al.\ 1995b, Fabian et al.\
1998).  The mean X-ray spectra of these objects appear to be
significantly harder than that of QSOs (Almaini et al.\ 1996,
Romero-Colmenero et al.\ 1996).  Recently, Schmidt et al.\ (1998) have
argued that confusion plays a significant role in the optical
identification of some of the faintest X-ray sources, and that some of
the emission-line galaxies identified previously are not the true
optical counterparts of the X-ray source.  Nevertheless, the hardness
of their X-ray spectra suggests that the true identifications will not
be standard broad emission-line QSOs.  Indeed, Roche et al.\
(1995) have cross-correlated faint galaxy positions with the
unresolved ROSAT X-ray background to demonstrate statistically that a
large fraction of the soft X-ray background must to due to faint $B
\sim 23\,$mag galaxies.
 
The {\it ASCA} mission has now yielded information on the nature of
the X-ray background at harder energies ($2-10\,$keV), albeit at a
poorer spatial resolution ($\sim 1\,$arcmin) than the {\it ROSAT}
mission. Although attempts to carry out optical identification
programs have been frustrated by the large {\it ASCA}
point-spread-function (PSF), Ohta et al.\ (1996) were able to identify
a high redshift, $z=0.9$, Type 2 AGN as a counterpart to a faint {\it
ASCA} source.  This source has an X-ray luminosity of $1 \times
10^{44}\,$erg$\,$s$^{-1}$, and was, until the observations reported on
in this paper, the only Type 2 AGN at high redshift ($z>0.1$)
to be identified on the basis of its hard $2-10\,$keV flux.
   
Many {\it ASCA} source identification programmes now rely on the
cross-correlation of the positions of sources detected in deep {\it
ASCA} data with those in deep {\it ROSAT} data obtained over the same
fields. This was the method we chose to employ in our programme of
faint {\it ASCA} source identifications.  Paper I in this series
(Georgantopoulos et al.\ 1997) reported on the source population and
the $2-10\,$keV $\log N - \log S$ in three deep {\it ASCA} fields
which had been cross-correlated with {\it ROSAT} data.  Paper II
(Boyle et al.\ 1998) presented the AGN $2-10\,$keV X-ray luminosity
function and the extrapolated AGN contribution to the $2-10\,$keV XRB.
We have since obtained data for a further two deep {\it ASCA} fields
and, in the course of the optical follow-up, identified an object with
narrow emission lines as the counterpart to one of the hardest {\it
ASCA} X-ray detections in our source list. In Section 2 we present the
optical and X-ray observations and in Section 3 we discuss the
properties of the object and present our conclusions.

\figure{1}{S}{0mm}{
\psfig{figure=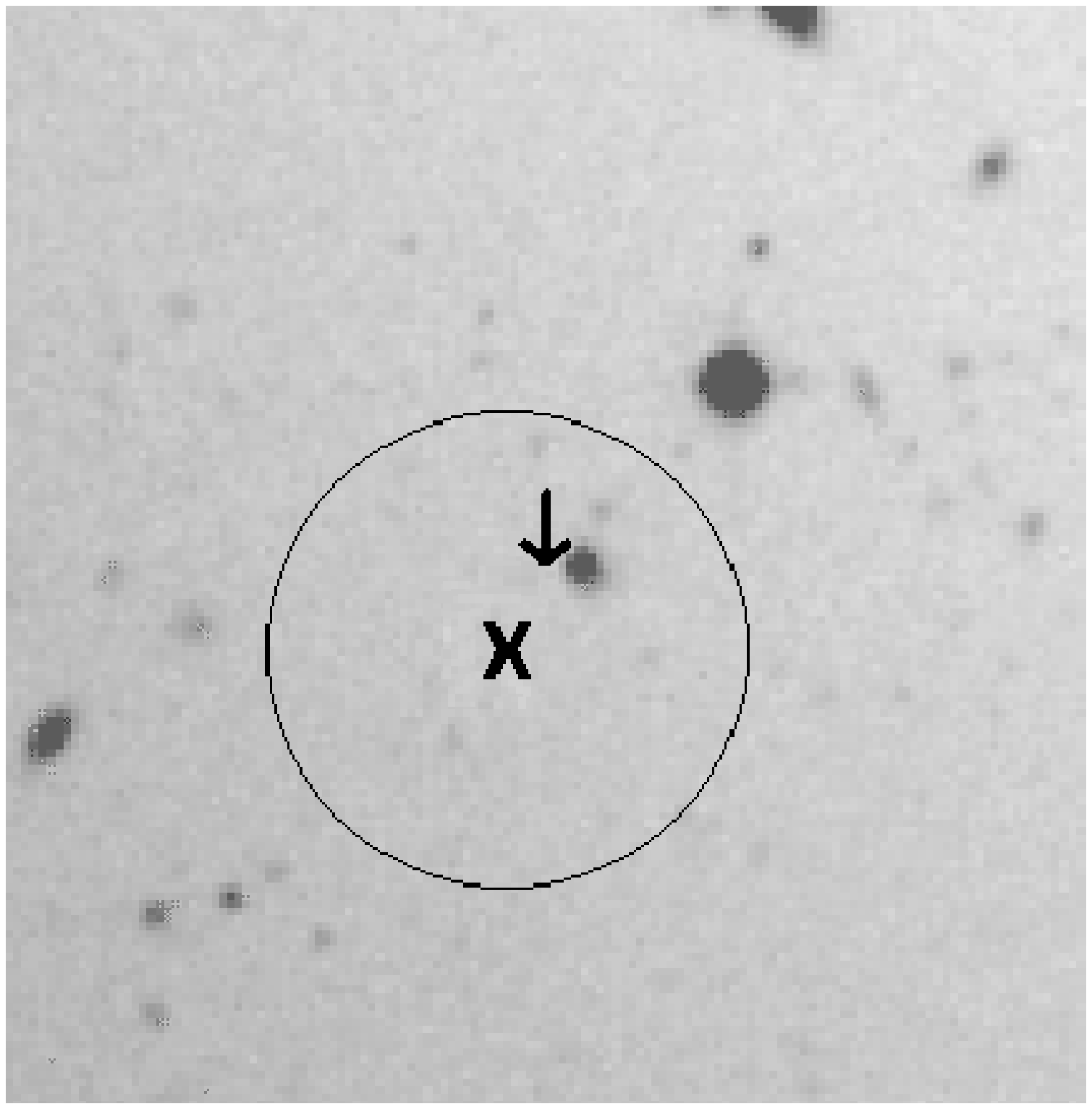,width=3.2in}\break
\noindent\bf Figure 1. \rm $1.5 \times 1.5$ arcmin region             
centred on the {\it ASCA} centroid (indicated by the arrow) for
AXJ0341.4--4453.  The {\it ROSAT} PSPC position is marked with a
cross, with an error circle of 20 arcsec radius.}

\section{Observations}\tx

\subsection{X-ray}\tx

We have obtained deep (100--200$\,$ksec) {\it ASCA} data in five fields
in which we had already carried out a deep ($S_{\rm 0.5-2keV} > 3\times
10^{-15}\,$erg$\,$s$^{-1}$cm$^{-2}$) {\it ROSAT} survey
(Georgantopoulos et al.\ 1996).  The deep
fields surveyed to date are centred in three locations; two pointings
in the South Galactic Pole (SGP) regions (00$^h$53$-$28$^{\circ}$),
two in the UKST 249 field (03$^h$44$-$45$^{\circ}$) and one in the
UKST F855 field (10$^h$46+00$^{\circ}$).  The X-ray images used for
source detection were obtained by a mosaic of the data from both Gas
Imaging Spectrometer (GIS) detectors, thus maximising the effective
exposure time. X-ray sources were then located using the Point Source
Search (PSS) algorithm of Allen (1992).  Full details of the X-ray
analysis (source detection, flux calculation) can be found in Paper I.

One source, centred at RA(2000) = $03^{\rm h}41^{\rm m}24.7^{\rm s}$,
Dec(2000)=$-44^{\circ} 53'01''$, was found to exhibit extreme X-ray
properties, with an exceptionally hard X-ray spectrum. The {\it ASCA}
count rate for the source (named AXJ0341.4--4453) based on the 
`cleaned' image is $1.7\pm0.46 \times 10^{-3}\,$count$\,$s$^{-1}$.
For an intrinsic energy spectral index of $\alpha=1$ and 
$n_{\rm H} = 4 \times 10^{22}$ cm$^{-2}$ (see below), we obtain 
a flux of $S_{\rm 2-10keV}=1.1\pm 0.27 
\times 10^{-13}\,$erg$\,$s$^{-1}$cm$^{-2}$.

AXJ0341.4--4453 is one of the brightest detections on the {\it ASCA}
field above $2\,$keV, but remains undetected in the softer $1-2\,$keV
band.  Although too faint to enable detailed spectral fitting, we note
however that the source shows an unusually large {\it ASCA} hardness
ratio $HR=0.91 \pm 0.24$, defined by: $HR={H-S\over{H+S}}$
\noindent
where $H$ and $S$ are the {\it ASCA} count rates in the {\it ASCA}
hard ($2-10\,$keV) and soft ($1-2\,$keV) bands respectively.  This
immediately suggests a hard, inverted X-ray spectrum or heavy
photo-electric absorption.  Given that there is no detection in the
soft band, it is more appropriate to give the lower limit on the
hardness ratio corresponding to the upper limit on the
soft count rate. We therefore estimated a lower bound on the hardness
ratio using the method of Kraft et al.\ (1991).  This gives a 99 per
cent lower limit of $HR > 0.34$, which still corresponds to a very
hard power law index of $\alpha < 0.1$.

AXJ0341.4--4453 was also found to have a faint $4.5\sigma$ ($S_{\rm
0.5-2keV} = 6\times 10^{-15}\,$erg$\,$s$^{-1}$cm$^{-2}$) {\it ROSAT}
source only $8\,$arcsec distant, well within the 90-arcsec correlation
length for {\it ASCA/ROSAT} positional coincidences
(Paper I). Assuming no X-ray variability between
the {\it ROSAT} and {\it ASCA} observation epochs, combining the {\it
ROSAT} and {\it ASCA} data yields a hardness ratio (as defined above)
of $HR=0.70\pm0.20$ or a spectral index of $\alpha =-0.8$ over the 
range $1-10\,$keV. 

\subsection{Optical}\tx

\figure{2}{D}{0mm}{
\psfig{figure=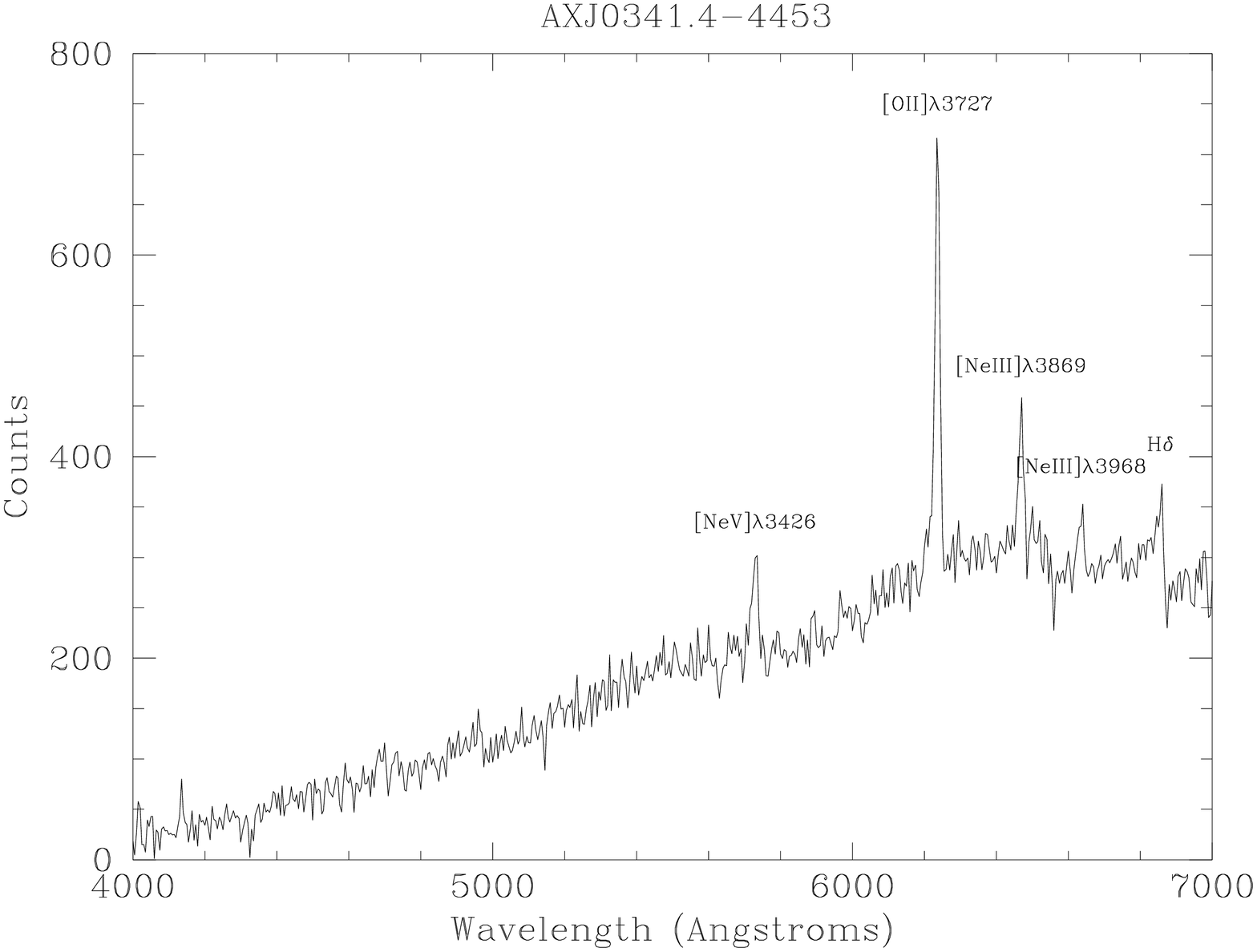,width=6.5in,angle=270}\break
\noindent\bf Figure 2. \rm LDSS optical spectrum for AXJ0341.4--4453}

The optical source closest to the {\it ROSAT} counterpart of
AXJ0341.4--4453 was a $B=21.4$ object only $11\,$arcsec from
the {\it ROSAT} position (see Fig.\ 1) and only  $3\,$arcsec
from the {\it ASCA} source position.  The optical position of 
the source is RA(2000)=$03^{\rm h}41^{\rm m}24.4^{\rm s}$, 
Dec(2000)=$-44^{\circ} 53'02''$.  Although
a spectroscopic identification already exists for many of the {\it
ROSAT} sources on this field (QSF1, Griffiths et al.\ 1995), the
optical counterpart to AXJ0341.4--4453 had not been observed
previously, because the {\it ROSAT} source was below the nominal X-ray
detection limit ($5\sigma$) adopted for completeness in the original
{\it ROSAT} survey.

The optical source was observed with the Low Dispersion Survey
Spectrograph (LDSS) at the Cassegrain focus of the Anglo-Australian
Telescope (AAT) on the night of 1997 September 6. 
The total integration time was 2400 seconds.  LDSS was operated
with a single `long-slit' aperture mask (1.7-arcsec slit width) and
medium resolution grism. A Tektronix CCD was used as the detector to
give an overall spectral resolution of 13\AA. Conditions were poor; 2
arcsec seeing and the observations were made through light cirrus.
Wavelength calibration was achieved using a copper-argon arc lamp.

The resulting optical spectrum is shown in Fig. 2.  The five strong
emission lines identified in the spectrum are listed in Table 1. The
mean emission line redshift obtained from these lines (excluding
H$\delta$, see below) is $z=0.6724\pm0.0002$. Based on this redshift,
the 2--10$\,$keV X-ray luminosity for AXJ0341.4--4453 is $1.8 \times
10^{44}\,$erg$\,$s$^{-1}$ ($q_0=0.5$, $H_0 = 
50\,$km$\,$s$^{-1}$Mpc$^{-1}$). 

The mean rest full width half maxima (FWHM) for the emission lines is
$900\pm150\,$km$\,$s$^{-1}$; marginally greater than the instrumental
profile of $\sim650\,$km$\,$s$^{-1}$.  The H$\delta$ emission line
appears to be cut on its red side by an absorption feature.  This is
consistent with the lower value of the emission line redshift obtained
for this line.  This absorption feature is also seen in the spectrum
of AXJ08494+4454, the $z=0.9$ Type 2 AGN at discovered by Ohta et al.\
(1996).

\table{1}{S}{\noindent\bf Table 1.\ \ \rm Emission Line Parameters for
AXJ0341.4--4453} 
{\tabskip=1em plus 2em minus .7em 
\halign to\hsize{#\hfil&\hfil#&\hfil#&\hfil#&
\hfil#\cr
\noalign{\medskip} 
\noalign{\hrule}
\noalign{\smallskip} 
Line&$\lambda_{\rm obs}$&\hfil $z$\hfil&$W_0$&FWHM$_0$\cr
&(\AA)&&(\AA)&(km$\,$s$^{-1}$)\cr
\noalign{\medskip} 
\noalign{\hrule}
\noalign{\smallskip} 
[NeV]$\lambda3426$&5729&0.6722&6.6&1140\cr
[OII]$\lambda3727$&6234&0.6727&14.7&740\cr
[NeIII]$\lambda3869$&6470&0.6723&5.7&830\cr
[NeIII]$\lambda3968$&6636&0.6724&2.4&900\cr
H$\delta$&6853&0.6706&3.6&870\cr
\noalign{\medskip}
\noalign{\hrule}
\noalign{\medskip}
}}

\section{Discussion}\tx

The strong [NeV] emission in the optical spectrum of AXJ0341.4--4453
indicates that this object is likely to be an AGN.  The lack of any
broad emission lines, particularly the broad MgII$\lambda$2798 line at
4673\AA\ further implies that this is an example of a high redshift
Seyfert 2 galaxy or Type 2 AGN.  This is confirmed by its high X-ray
luminosity ($1.8 \times 10^{44}\,$erg$\,$s$^{-1}$), unlikely to
originate from a star-forming galaxy. The optical spectrum bears many
similarities to that obtained for the $z=0.9$ Type 2 AGN observed by
Ohta et al.\ (1996), and other Type 2 AGN identified by {\it ROSAT} in
Lockman hole field (Schmidt et al.\ 1998).

The identification of such sources as bona fide Type 2 AGN has
recently been called into question by Halpern and co-workers in a
series of papers, see Halpern, Eracleous \& Forster (1998) and
references therein.  They have demonstrated that many of the sources
initially identified as Type 2 AGN either exhibit broad Balmer lines
and are better classified as high-redshift Seyfert 1.8 or Seyfert 1.9
galaxies, or are members of the class of narrow-line Seyfert 1
galaxies (Brandt et al. 1994), with steep (soft) X-ray spectral
indices.  Given its hard X-ray spectrum, we can immediately rule out
AXJ0341.4--4453 as a narrow-line Seyfert 1.

While it is impossible to rule out a Seyfert 1.8/1.9 classification on
the basis of the only Balmer line (H$\delta$) seen in the spectrum of 
AXJ0341.4--4453, we can place a useful 3$\sigma$
upper limit of 10\AA\ on the rest equivalent width of any broad
(5000km$\,$s$^{-1}$ FWHM) MgII$\lambda 2798$ emission line.  This is
well below the typical equivalent width of broad MgII seen in the
objects observed to date by Halpern et al.\ (e.g. $W_0 = 47$\AA\ for
1E$\,0449,4-1823$). Note there may be a weak ($W_0\sim5$\AA), narrow
(FWHM$< 2000\,$km$\,$s$^{-1}$) emission feature at 
$\lambda \sim 4690$\AA\
in the spectrum of AX$\,$J0341.4--4453.  However, the association of
this weak feature with MgII$\lambda 2798$ is uncertain given that it is
redshifted $650\,$km$\,$s$^{-1}$ with respect to the narrow emission
lines of [OII], [NeIII] and [NeV], and by over $1000\,$km$\,$s$^{-1}$
from the H$\delta$ emission line.

Indeed, the classification of these objects as Seyfert 1.8/1.9 or
Seyfert 2 may well be a largely taxonomic issue.  More fundamental to
the nature of the AXJ0341.4--4453 is the fact that it exhibits a hard
X-ray spectrum.  If we assume that absorption causes the flattening of
the hardness ratios, the {\it ASCA} $1-10\,$keV hardness ratio
($HR>0.34$) yields a 99 per cent confidence lower limit on the
photo-electric absorption of $n_{\rm H} > 4 \times 10^{22}$ cm$^{-2}$,
for an intrinsic energy spectral index $\alpha=1$.  The combined {\it
ROSAT/ASCA} hardness ratio ($HR=0.70$) yields an absorption of $n_{\rm
H} \sim 10^{23}$ cm$^{-2}$, for $\alpha=1$.  The dust associated with
such a column would be more than enough to extinguish the broad line
region in the optical waveband.  We conclude that this is an example
of an AGN obscured by gas and dust, of the type widely postulated to
account for the hard X-ray background (Madau et al.\ 1994, Comastri et
al.\ 1995).

AXJ0341.4--4453 also exhibits the highest value $\alpha_{\rm OX}$ for
any object in our survey, $\alpha_{\rm OX}=1.8$ ($\alpha_{\rm
OX}=-\log(L_{\rm 2keV}/L_{\rm 2500\AA})/2.605$).  This may be compared
to mean value of $\alpha_{\rm OX}=1.45 \pm 0.10$ for the survey sources
as a whole. This is consistent with significant absorption of the 2keV
X-ray flux by a large HI column and is typical of the values obtained
for Seyfert 2 galaxies at low redshift (see e.g. Rush et al.\ 1996).

Examples of obscured AGN have also been found at low redshift
($z<0.1$) from {\it ASCA} observations. Akiyama et al.\ (1998) report
on the discovery of a type 2 Seyfert galaxy, AX$\,$J131501+3141, at
$z=0.072$.  This object exhibits a similar absorption column to that
obtained for AX$\,$J0341.4--4453 ($n_{\rm H} \sim 6 \times 10^{22}$
cm$^{-2}$) and is the hardest X-ray source in the {\it ASCA} large sky
survey.  Another source AX$\,$J1749+684, identified by Iwasawa et al.\
(1997), may also be an example of a low redshift obscured AGN, but
with a lower absorbing column $n_{\rm H} \sim 8 \times 10^{21}$ 
cm$^{-2}$. 

If obscured AGN such as AXJ0341.4--4453 are to explain the
hard XRB then they should emerge in significant numbers in
forthcoming, deeper X-ray surveys (e.g. with {\it AXAF} or {\it XMM}).
They should also turn up in the latest mid-
and far-infrared deep surveys, where the energy absorbed by the
obscuring gas and dust will be re-radiated (Granato et al.\ 1997).
At mid-infrared wavelengths the European Large Area Infrared Space
Observatory Survey (ELAIS, Oliver et al.\ 1997) will reach
unprecedented sensitivities, and large numbers of obscured AGN should
be identified (Barcons et al.\ 1995).  For obscured QSOs at high
redshift, the expected dust emission peak at $\lambda \sim 60-100
\mu$m is redshifted into the sub-millimetre region. Deep surveys with
the new Sub-mm Common User Bolometer Array (SCUBA) at the James Clerk
Maxwell Telescope in Hawaii therefore present an ideal opportunity to
detect these obscured objects, if they are truly the explanation for
the origin of the hard XRB.

\section*{Acknowledgements}\tx

\noindent  Optical spectroscopy was carried out with the 
Anglo-Australian Telescope, based on initial identifications of
candidates obtained from APM measurements of UK Schmidt telescope
plates. OA, GCS and KFG acknowledge PPARC for support.

\section*{References}
\bibitem
Akiyama M., et al.\ 1998, ApJ, in press (astro-ph/9801173)
\bibitem 
Allen D.J., 1992, Starlink ASTERIX User Note 004
\bibitem
Almaini O., Boyle B.J., Griffiths R.E., Shanks T., Stewart G.C.,
Georgantopoulos I., 1995, MNRAS, 277, L31
\bibitem
Almaini O., Shanks T., Boyle B.J., Griffiths R.E.,
Roche N., Stewart G.C., Georgantopoulos, I., 1996, MNRAS, 282, 295
\bibitem
Barcons X., Franceschini A., De Zotti G., Danese L., Miyaji T., 1995, 
ApJ, 455, 480
\bibitem
Boyle B.J., Georgantopoulos I., Blair A., Stewart G.C., Griffiths R.E.,Shanks T., Gunn K.F., Almaini O., 1998, MNRAS, in press
\bibitem
Boyle B.J., Griffiths R.E., Shanks T., Stewart G.C., 
Georgantopoulos I., 1994, MNRAS, 271, 639
\bibitem
Boyle B.J., McMahon R.G., Wilkes B.J., Elvis M., 1995a, MNRAS
272, 462
\bibitem
Boyle B.J., McMahon R.G., Wilkes B.J., Elvis M., 1995b, MNRAS, 276, 315
\bibitem
Brandt W.N., Fabian A.C., Nandra K., Reynolds C.S., Brinkmann W., 
MNRAS, 271, 958
\bibitem
Comastri A., Setti G., Zamorani G., Hasinger G., 1995,
A\&A, 296, 1
\bibitem
Fabian A.C., Barcons X., Almaini O., Iwasawa K., 1998, MNRAS submitted
\bibitem
Georgantopoulos I., Stewart G.C., Blair A.J., Shanks T., 
Griffiths R.E., Boyle B.J., Almaini O., Roche N., 1997, MNRAS,
291, 203 (Paper I)
\bibitem
Georgantopoulos I., Stewart G.C., Shanks T., Boyle B.J.,
Griffiths R.E., 1996, MNRAS, 280, 276
\bibitem
Griffiths R.E., Della Ceca R., Georgantopoulos I, Boyle B.J, 
Stewart G.C., Shanks T., Fruscione A., 1996, MNRAS, 281, 71
\bibitem
Griffiths R.E., Georgantopoulos I, Boyle B.J, 
Stewart G.C., Shanks T., Della Ceca R., 1995, MNRAS, 275, 77
\bibitem
Granato G.L., Danese L., Franceschini A., 1997, ApJ, 486, 147
\bibitem
Halpern J.P., Eracleous M., Forster K., 1998, ApJ, in press
\bibitem
Iwasawa K., Fabian A.C., Bradt W.N., Crawford C.S., Almaini O., 1997, MNRAS, 291, L17
\bibitem
Kraft R.P., Burrows D.N., Nousek J.A., 1991, ApJ, 374, 344
\bibitem
Madau P., Ghisellini G., Fabian A.C., 1994, MNRAS, 270, 17
\bibitem
Ohta K., Yamada T., Nakanishi K., Ogasaka Y., Kii T., Hayashida K., 
1996,  ApJ, 458, L57
\bibitem
Oliver S.J. et al 1997, in McLean B. et al. eds, Proc. IAU
Symp. 179, New Horizons from Multi-Wavelength Sky Surveys, Kluwer,
Dordrecht, in press
\bibitem
Roche N., Shanks T., Georgontopoulos I., Stewart G.C., Boyle
B.J., Griffiths R.E., 1995, MNRAS, 273, L15
\bibitem
Romero-Colmenero E., Branduardi-Raymont G., Carrera F.J., Jones L.R.,
Mason K.O., McHardy I.M., Mittaz J.P.D., 1996, MNRAS, 282, 94
\bibitem
Rush B., Malkan M.A., Fink H.H., Voges W., 1996, ApJ, 471, 190
\bibitem
Schmidt M. et al.\ 1998, A.\&A., 329, 495
\bibitem
Shanks T., Geogantopoulos I., Stewart G.C., Pounds K.A., Boyle B.J.,
Griffiths R.E., 1991, Nat, 353, 315
\bibitem
\bye